\documentclass[conference]{IEEEtran}
\IEEEoverridecommandlockouts
\usepackage{cite}
\usepackage{amsmath,amssymb,amsfonts}
\usepackage{algorithm}
\usepackage{algorithmic}
\usepackage{graphicx}
\usepackage{textcomp}
\usepackage{xcolor}
\usepackage{subcaption}
\usepackage{multirow}
\usepackage{array}
\usepackage{hyperref}
\usepackage[symbol]{footmisc}
\usepackage{balance}

\usepackage{authblk}

\def\BibTeX{{\rm B\kern-.05em{\sc i\kern-.025em b}\kern-.08em
    T\kern-.1667em\lower.7ex\hbox{E}\kern-.125emX}}
\begin{document}

\title{Travelling Salesman Problem: \\ Parallel Implementations \& Analysis

}

\author[1]{Amey Gohil}
\author[1]{Manan Tayal}
\author[1]{Tezan Sahu}
\author[1]{Vyankatesh Sawalpurkar}
\affil[1]{BTech(2021), Indian Institute of Technology Bombay, Mumbai, India}

\maketitle

\begin{abstract}
The Traveling Salesman Problem (often called TSP) is a classic algorithmic problem in the field of computer science and operations research. It is an NP-Hard problem focused on optimization. TSP has several applications even in its purest formulation, such as planning, logistics, and the manufacture of microchips; and can be slightly modified to appear as a sub-problem in many areas, such as DNA sequencing. In this paper, a study on parallelization of the \emph{Brute Force approach} (under several paradigms) of the Travelling Salesman Problem is presented. Detailed timing studies for the serial and various parallel implementations of the Travelling Salesman Problem have also been illustrated.
\end{abstract}


\section{Introduction}

The Travelling Salesman Problem (TSP) is the challenge of finding the shortest yet most efficient route for a person to take given a list of specific destinations along with the cost of travelling between each pair of destinations.

Stated formally, given a set of $N$ cities and distances between every pair of cities, the problem is to find the shortest possible route that visits every city exactly once and returns to the starting point. The problem is an NP-Hard problem that is, no polynomial-time solution exists for this problem. The brute force solution for the problem is to consider $city_1$ as a starting city and then generate all the permutations of the remaining $N-1$ cities and return the permutation with the minimum cost. The time complexity for this solution is $O(N!)$. Another solution using the Dynamic programming paradigm exists with the time complexity $O(N^{2}2^{N})$ which is much less than $O(N!)$ but it has exponential space complexity so impractical to implement.

The goal of this study is to parallelize the Brute Force algorithm for solving TSP using a variety of paradigms (using OpenMP, MPI \& CUDA), and to critically compare and analyse the differences in their performance.

\section{Serial Implementation of Brute Force TSP}

Representing a TSP tour is rather simple. It is just a permutation of the cities, with the added restriction that the first city must always be $city_1$. Once we realize that each of the tours is a permutation, then a brute-force algorithm that is guaranteed to always solve the TSP becomes evident: Examine all possible permutations of cities, and keep the one that is shortest. The pseudo-code for this approach has been outlined in Algorithm \ref{alg:brute_force_tsp}.

The number of permutations made using $city_1, city_2, ..., city_N$ is given by $N!$ Since we only want those permutations with $city_1$ as the first destination, we are left with $(N-1)!$ permutations to explore. Hence, iterating through the permutations takes $O((N-1)!)$ time. 

\vspace{-5pt}
\begin{algorithm}[!ht]
    \caption{Brute Force Serial TSP}
    \label{alg:brute_force_tsp}
\begin{algorithmic}
\STATE {\bfseries Input:} $city\_list$, $cost\_matrix$

\STATE $optimal\_cost \gets \infty$
\STATE $optimal\_path\gets null$
\STATE $city\_list\gets list\ of\ cities\ excluding\ city_1$
\WHILE {$next\_permutation(city\_list)$}
        \STATE $temp\_cost\gets get\_path\_cost(city\_list, cost\_matrix)$ 
        \COMMENT{Cost of travelling cities in order of cities in $city\_list$}
        
        \IF{$temp\_cost < optimal\_cost$}
        \STATE $optimal\_cost\gets temp\_cost\ $
        \STATE $optimal\_path\gets ${$city\_list$}
        \COMMENT{with $city_1$ appended at start \& end}
        \ENDIF

\ENDWHILE

\STATE {\bfseries Output:} $optimal\_path$, $optimal\_cost$
\end{algorithmic}
\end{algorithm}

\vspace{-5pt}

Now, calculating the cost for $N$ cities in the path for each permutation requires traversal of an array of length $N$, which takes $O(N)$ time. Hence, this brute force algorithm takes $O((N-1)! \times N) = O(N!)$ time.

Algorithm \ref{alg:brute_force_tsp} shows the use of the $next\_permutation(...)$ function that takes in a list of destinations, i.e.,  $city\_list$ \& checks if there exists a next lexicographic permutation for the given permutation of destinations. If it does exist, modify $city\_list$ to store this next permutation, so that it may be used to calculate the path cost. The $get\_path\_cost(...)$ function calculates the cost of travelling through the cities based on the order mentioned in $city\_list$, \& finding the individual trip costs from the $cost\_matrix$. At the end, we would have exhausted all possible $(N-1)!$ permutations \& arrive at the \emph{optimal path} with the \emph{least path cost}.

This serial implementation of Brute Force TSP can be found in \texttt{tsp.cpp} file.

\section{Motivation to Parallelize Brute Force TSP}

We notice that all the major computations required for finding the optimal path in this brute force approach need to be done for every permutation. Clearly, parallelizing the iteration over these permutation of destinations makes this problem almost \emph{embarrassingly parallelizable} under brute force search.

Thus, our project involves the implementation of this brute force TSP algorithm under different paradigms using: 

\begin{itemize}
    \item OpenMP (Shared Memory Processing)
    \item MPI (Message Passing Interface)
    \item CUDA (for NVIDIA GPUs)
\end{itemize}

Further, our analysis focuses on the  comparison between the time taken \footnote[1]{for each reading, we report the average of 5 runs} for the serial algorithm \& its various parallel counterparts, while also trying to explain the parallelizability of the algorithm.

\section{System Specifications for Experiments}

All the code execution \& timing studies have been performed on \emph{Param Sanganak}\footnote[2]{A Supercomputer Facilty at IIT Kanpur}. Tables \ref{tab:hw_specs} \& \ref{tab:sw_specs} summarize the specifications of the execution environment.

\begin{table}[h]
    \centering
    \setlength\extrarowheight{2pt}
    \begin{tabular}{|c|c|}\hline
         \textbf{Architecture} & x86\_64 \\
         \textbf{CPU op-mode(s)} & 32-bit, 64-bit \\
         \textbf{Byte Order} & Little Endian \\
         \textbf{CPU(s)} & 40 \\
         \textbf{Thread(s) per core} & 1 \\
         \textbf{Core(s) per socket} & 20 \\
         \textbf{Socket(s)} & 2 \\
        %  \textbf{Vendor ID} & GenuineIntel \\
         \multirow{2}{*}{\textbf{Model Name}} & Intel(R) Xeon(R) Gold\\&  6248 CPU @ 2.50GHz\\
         \textbf{CPU MHz} & 999.908 \\ 
         \textbf{GPU Model} & Tesla V100-SXM2-16GB \\
         \textbf{GPU Bus Type} & PCIe \\
         \textbf{GPU DMA Size} & 47 bits \\ \hline
    \end{tabular}
    \caption{Hardware Specifications on Param Sanganak}
    \label{tab:hw_specs}
\end{table}

\begin{table}[h]
\centering
\setlength\extrarowheight{2pt}
\begin{tabular}{|c|c|}\hline
     \textbf{Operating System} & CentOS 7.6 \\
     \textbf{\texttt{g++} Version} & 4.8.5 \\
     \textbf{OpenMP Version} & 3.1 \\
     \textbf{MPI Version} & 4.0.2rc3 \\ 
     \textbf{CUDA Version} & 11.1\\\hline
\end{tabular}
\caption{Software Specifications on Param Sanganak}
\label{tab:sw_specs}
\end{table}

\section{OpenMP (Shared Memory Processing)}

\subsection{Approach for Parallelization}
For parallelizing the brute force approach for TSP using OpenMP, all the different permutations of cities are consider and are divided equally (in case number of threads divide number of permutations exactly) among different sections based on threads, however if number of threads does not exactly divide number of permutations, the first r (remainder on division) threads are given one extra permutation to compute. Now, to maintain continuity between two consecutive threads, the initial arrangements of cities are given to each section based on thread IDs and then consequent arrangements are computed within the thread itself. Each thread now calculates its respective optimal cost and path, by looping between the starting and ending permutation (for each thread) and then comparing the optimal values and path of all the threads. This implementation can be found in \texttt{tsp\_omp.cpp} file.

\subsection{Timing Analysis}
\begin{figure}[h]
    \centering
    \includegraphics[width=0.8\linewidth]{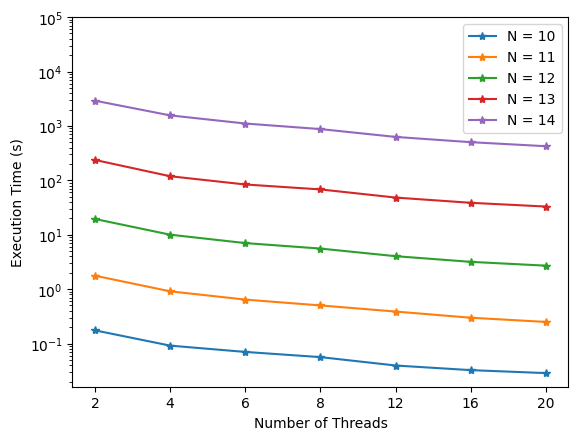}
    \caption{Variation of execution times with number of OpenMP threads for different problem sizes ($N = $ Number of cities)}
    \label{fig:openmp_time}
\end{figure}

\begin{figure}[h]
    \centering
    \includegraphics[width=0.8\linewidth]{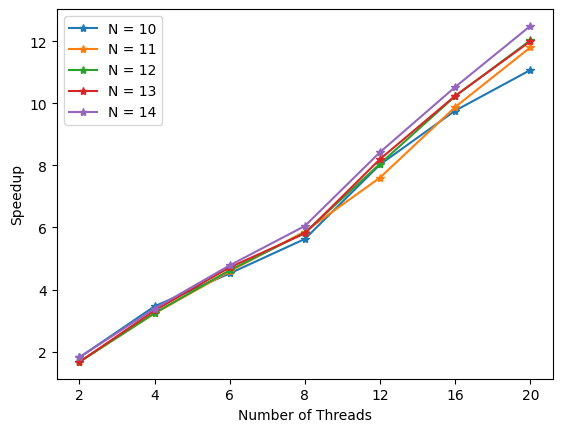}
    \caption{Variation of speedup achieved with number of OpenMP threads for different problem sizes ($N = $ Number of cities)}
    \label{fig:openmp_speedup}
\end{figure}

Figures \ref{fig:openmp_time} \& \ref{fig:openmp_speedup} illustrate the time taken for execution and speedup achieved for the OpenMP parallelization approach for different number of threads and number of cities (N). 

\begin{itemize}
    \item Execution time decreases with increase in \# of threads.
    % \item Execution time increases with increase in \# of cities ($N$).
    \item The difference between the execution time for two consecutive cities is large because it depends on $N!$.
    \item Speedup increases with increase in the number of threads, while the trend remains similar across all $N's$.
\end{itemize}

\section{MPI (Message Passing Interface)}

\subsection{Approach for Parallelization}

For parallelizing the brute force approach for TSP using MPI (Message Passing Interface), the permutations of cities are equally divided among available Parallel Environments (PEs), while taking care of the case where equal distribution is not possible. The data of edge weights is distributed across all PEs, followed by synchronization. Each PE is given a starting permutation index and an ending permutation index. Calculation of these permutation indices are done so as to take care of the cases when equal distribution of permutations is not possible, and therefore, remainder work is further divided across PEs. Each PE now calculates its respective optimal cost and path, by looping through the starting to ending permutation, following the same procedure as in serial algorithm. For each PE, we have their individual respective variables to calculate optimal path and optimal value from among the permutations assigned to them. Now, all PEs are synchronized and optimal costs and paths of all PEs are compared in the \emph{Master PE}. This implementation can be found in \texttt{tsp\_mpi.cpp} file.

\begin{figure}[h]
    \centering
    \includegraphics[width=0.8\linewidth]{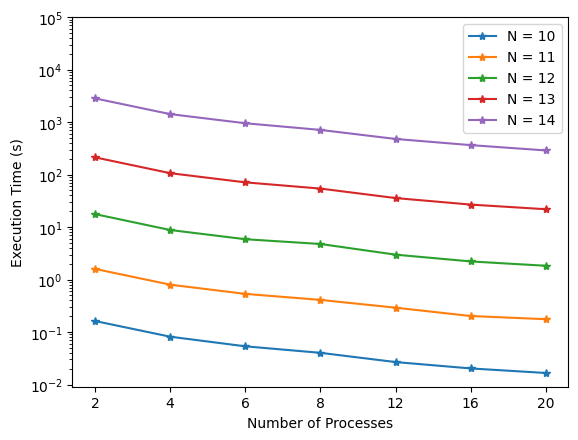}
    \caption{Variation of execution times with number of MPI processes for different problem sizes ($N = $ Number of cities)}
    \label{fig:mpi_time}
\end{figure}

\begin{figure}[h]
    \centering
    \includegraphics[width=0.8\linewidth]{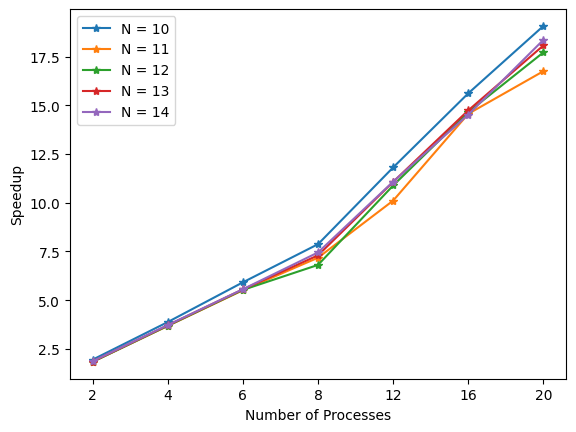}
    \caption{Variation of speedup achieved with number of MPI processes for different problem sizes ($N = $ Number of cities)}
    \label{fig:mpi_speedup}
\end{figure}

\subsection{Timing Analysis}

Figures \ref{fig:mpi_time} \& \ref{fig:mpi_speedup} illustrate the time taken for execution and speedup achieved for the MPI parallelization approach for different number of Parallel Environments (PEs) and number of cities ($N$).

\begin{itemize}
    \item Execution time decreases with increase in \# of PEs.
    % \item Execution time increases with increase in \# of cities($N$).
    \item The difference between the execution time for two consecutive cities is large because it depends on $N!$.
    \item Speedup increases with increase in the number of PEs, while the trend remains similar across all $N's$.
\end{itemize}

\section{Comparative Study of OpenMP \& MPI}

Based on the Timing Analysis of the OpenMP \& MPI implementations, we conclude that the speedups achieved through Message Passing (i.e., MPI) are greater than those achieved through Shared Memory (i.e., OpenMP). This is apparent from the fact that using 20 PEs (in MPI) results in speedups of  $\sim18$, while using 20 threads (in OpenMP) results in speedups of merely $\sim12$.

\begin{table}[!h]
\centering
\setlength\extrarowheight{2pt}
\begin{tabular}{c|c|ccccc|}
\cline{3-7}
\multicolumn{2}{c|}{\multirow{2}{*}{}} & \multicolumn{5}{c|}{\textbf{No. of Threads (p)}}                 \\ \cline{3-7} 
\multicolumn{2}{c|}{}                  & \textbf{2} & \textbf{4} & \textbf{8} & \textbf{16} & \textbf{20} \\ \hline
\multicolumn{1}{|c|}{\multirow{5}{*}{\textbf{N}}} & \textbf{10} & 0.907 & 0.865 & 0.704 & 0.610 & 0.554 \\
\multicolumn{1}{|c|}{}  & \textbf{11}  & 0.834      & 0.808      & 0.734      & 0.618       & 0.590       \\
\multicolumn{1}{|c|}{}  & \textbf{12}  & 0.834      & 0.810      & 0.729      & 0.639       & 0.601       \\
\multicolumn{1}{|c|}{}  & \textbf{13}  & 0.834      & 0.831      & 0.726      & 0.640       & 0.600       \\
\multicolumn{1}{|c|}{}  & \textbf{14}  & 0.909      & 0.848      & 0.750      & 0.658       & 0.618       \\ \hline
\end{tabular}
\caption{Efficiency ($\eta$) of the OpenMP implementation for various $N$ (problem size) \& $p$ (no. of OpenMP threads)}
\label{tab:eff_openmp}
\end{table}

\begin{table}[!h]
\centering
\setlength\extrarowheight{2pt}
\begin{tabular}{c|c|ccccc|}
\cline{3-7}
\multicolumn{2}{c|}{\multirow{2}{*}{}} & \multicolumn{5}{c|}{\textbf{No. of PEs (p)}}                 \\ \cline{3-7} 
\multicolumn{2}{c|}{}                  & \textbf{2} & \textbf{4} & \textbf{8} & \textbf{16} & \textbf{20} \\ \hline
\multicolumn{1}{|c|}{\multirow{5}{*}{\textbf{N}}} & \textbf{10} & 0.975 & 0.970 & 0.984 & 0.976 & 0.953 \\
\multicolumn{1}{|c|}{}  & \textbf{11}  & 0.922      & 0.920      & 0.898      & 0.910       & 0.838       \\
\multicolumn{1}{|c|}{}  & \textbf{12}  & 0.918      & 0.922      & 0.851      & 0.916       & 0.886       \\
\multicolumn{1}{|c|}{}  & \textbf{13}  & 0.927      & 0.929      & 0.913      & 0.921       & 0.904       \\
\multicolumn{1}{|c|}{}  & \textbf{14}  & 0.934      & 0.931      & 0.931      & 0.908       & 0.918       \\ \hline
\end{tabular}
\caption{Efficiency ($\eta$) of the MPI implementation for various $N$ (problem size) \& $p$ (no. of MPI Processes)}
\label{tab:eff_mpi}
\end{table}

\vspace{-5pt}

This is also evident by comparing the efficiencies of OpenMP \& MPI implementations shown in Tables \ref{tab:eff_openmp} \& \ref{tab:eff_mpi}. Clearly, the efficiency for MPI code is always greater than its OpenMP counterpart. This is potentially due to extra overheads of spawning \& maintaining OpenMP threads compared to the inter-process communications in MPI.

We also calculate the \emph{Karp-Flatt Metric}, or \emph{experimentally determined serial fraction $e(N, p)$ of parallel  computation} for both these implementations. Given the speedup $\Psi(N, p)$ using $p$ processors, $e(N, p)$ is determined as follows:
$$e(N, p) = \frac{\frac{1}{\Psi} - \frac{1}{p}}{1 - \frac{1}{p}}$$

\vspace{-10pt}

\begin{table}[!h]
\centering
\setlength\extrarowheight{2pt}
\begin{tabular}{c|c|ccccc|}
\cline{3-7}
\multicolumn{2}{c|}{\multirow{2}{*}{}} & \multicolumn{5}{c|}{\textbf{No. of Threads (p)}}                 \\ \cline{3-7} 
\multicolumn{2}{c|}{}                  & \textbf{2} & \textbf{4} & \textbf{8} & \textbf{16} & \textbf{20} \\ \hline
\multicolumn{1}{|c|}{\multirow{5}{*}{\textbf{N}}} & \textbf{10} & 0.103 & 0.052 & 0.060 & 0.043 & 0.042 \\
\multicolumn{1}{|c|}{}  & \textbf{11}  & 0.199      & 0.079      & 0.052      & 0.041       & 0.037       \\
\multicolumn{1}{|c|}{}  & \textbf{12}  & 0.199      & 0.078      & 0.053      & 0.038       & 0.035       \\
\multicolumn{1}{|c|}{}  & \textbf{13}  & 0.198      & 0.068      & 0.054      & 0.037       & 0.035       \\
\multicolumn{1}{|c|}{}  & \textbf{14}  & 0.100      & 0.060      & 0.048      & 0.035       & 0.033       \\ \hline
\end{tabular}
\caption{$e(N, p)$ of the OpenMP implementation for various $N$ (problem size) \& $p$ (no. of OpenMP threads)}
\label{tab:e_openmp}
\end{table}

\begin{table}[!h]
\centering
\setlength\extrarowheight{2pt}
\begin{tabular}{c|c|ccccc|}
\cline{3-7}
\multicolumn{2}{c|}{\multirow{2}{*}{}} & \multicolumn{5}{c|}{\textbf{No. of PEs (p)}}                 \\ \cline{3-7} 
\multicolumn{2}{c|}{}                  & \textbf{2} & \textbf{4} & \textbf{8} & \textbf{16} & \textbf{20} \\ \hline
\multicolumn{1}{|c|}{\multirow{5}{*}{\textbf{N}}} & \textbf{10} & 0.025 & 0.010 & 0.002 & 0.002 & 0.003\\
\multicolumn{1}{|c|}{}  & \textbf{11}  & 0.085      & 0.029      & 0.016      & 0.007       & 0.010       \\
\multicolumn{1}{|c|}{}  & \textbf{12}  & 0.089      & 0.028      & 0.025      & 0.006       & 0.007       \\
\multicolumn{1}{|c|}{}  & \textbf{13}  & 0.078      & 0.026      & 0.014      & 0.006       & 0.006       \\
\multicolumn{1}{|c|}{}  & \textbf{14}  & 0.070      & 0.025      & 0.011      & 0.007       & 0.005       \\ \hline
\end{tabular}
\caption{$e(N, p)$ of the MPI implementation for various $N$ (problem size) \& $p$ (no. of MPI Processes)}
\label{tab:e_mpi}
\end{table}

Tables \ref{tab:e_openmp} \& \ref{tab:e_mpi} show $e(N, p)$ for OpenMP \& MPI respectively. The following inferences can be made:

\vspace{-5pt}
\begin{itemize}
    \item For a given $N$, $e(N, p)$ usually decreases or remains constant with increase in $p$, indicating that the brute force TSP algorithm is rather embarrassingly parallel.
    \item Comparing the corresponding $e(N, p)$ values for OpenMP \& MPI, highlights that OpenMP code contains higher fraction of serial component, leading to lower speedups.
\end{itemize}

\vspace{-5pt}

\section{Hybrid Approach (Combining MPI \& OpenMP)}

\subsection{Approach for Parallelization}
After comparing the MPI \& OpenMP approaches, we implemented a hybrid approach by combining the two \& analysed its performance. This involved the spawning of multiple OpenMP threads ($num\_threads$) in each of the multiple MPI processes ($num\_PEs$) such that the permutations were divided equally among the $num\_PEs \times num\_threads$ parallel elements (while addressing the cases of unequal division). The code for this approach has been implemented in \texttt{tsp\_hybrid.cpp}.

\vspace{-5pt}

\subsection{Timing Analysis}

Keeping the total number of parallel elements same, increasing the number of MPI processes \& decreasing the number of OpenMP threads reduces the execution time. This can be explained by the larger overheads of maintaining OpenMP threads compared to the MPI communication overhead. A glimpse of this trend can be seen in Table \ref{tab:hybrid}.

However, using lesser number of PEs (with same total parallel elements) proves to be worse even compared to OpenMP, perhaps due to the higher costs of maintaining threads \& their synchronization across only a small number of PEs. Once the number of PEs becomes large enough, thread maintenance costs are overcome by MPI communication.

% As expected, for any given problem size $N$, the execution times for the hybrid approach with same number of total parallel elements were between those for OpenMP (only threads) \& MPI (only PEs). Moreover, increasing the number of MPI processes \& decreasing the number of OpenMP threads causes a reduction in the execution time. This can again be attributed to the larger overheads of maintaining OpenMP threads compared to the MPI communication overhead.

\begin{table}[!h]
\centering
\setlength\extrarowheight{2pt}
\begin{tabular}{|c|c|c|}
\hline
\multicolumn{2}{|c|}{\textbf{Parallelization Paradigm}}      & \textbf{Execution Time (s)} \\ \hline
\multicolumn{2}{|c|}{OpenMP (20 threads)}    & 32.934                      \\ \hline
\multicolumn{2}{|c|}{MPI (20 PEs)}           & 21.866                      \\ \hline
\multirow{4}{*}{Hybrid} & 2 PEs $\times$ 10 threads & 124.736                     \\
                        & 4 PEs $\times$  5 threads  & 62.219                      \\
                        & 5 PEs $\times$  4 threads  & 50.492                      \\
                        & 10 PEs $\times$  2 threads & 25.822                      \\ \hline

\end{tabular}
\caption{Comparison of hybrid code execution times against OpenMP \& MPI codes with same number of parallel elements ($20$) for $N = 13$}
\label{tab:hybrid}
\end{table}

\vspace{-5pt}

\section{CUDA (NVIDIA GPUs)}

\subsection{Approach for Parallelization}
For parallelizing the brute force approach for TSP using CUDA, we leverage both: \emph{blocks} \& \emph{threads}. For our analysis, we use \emph{50 blocks}, each with \emph{1024 threads}. We divide the permutations of cities among the threads in all the blocks. Each thread in a block calculates the minimum cost path from the assigned permutations. After synchronizing the threads in each separate block, one thread \textit{(with Thread ID = 0)} computes the optimal cost \& path for the block, and stores it in a shared global data structure that can be accessed by both GPU and CPU. Once the GPU finishes execution, the host calculates the optimal path by iterating over the minimum cost path of each block stored in the previously mentioned data structure. This implementation can be found in \texttt{tsp\_cuda.cu} file.

\subsection{Timing Analysis}

Table \ref{tab:cuda} summarizes the  time  taken  for  execution  and speedup achieved by leveraging the power of NVIDIA GPU.
\begin{table}[!h]
\centering
\setlength\extrarowheight{2pt}
\begin{tabular}{|c|cc|c|}
\hline
\multirow{2}{*}{\textbf{N}} & \multicolumn{2}{c|}{\textbf{Time (s)}}                          & \multirow{2}{*}{\textbf{Speedup}} \\ \cline{2-3}
                            & \multicolumn{1}{c|}{\textbf{Serial Code}} & \textbf{CUDA Code} &                                   \\ \hline
8  & \multicolumn{1}{c|}{0.004}     & 0.017 & 0.264x    \\
9  & \multicolumn{1}{c|}{0.065}     & 0.042 & 1.537x   \\
10 & \multicolumn{1}{c|}{0.316}     & 0.020 & 15.725x   \\
11 & \multicolumn{1}{c|}{2.9416}    & 0.057 & 51.278x   \\
12 & \multicolumn{1}{c|}{32.440}    & 0.048 & 673.260x  \\
13 & \multicolumn{1}{c|}{395.165}  & 0.086 & 4557.187x \\
14 & \multicolumn{1}{c|}{5289.317} & 0.822 & 6427.022x \\ 
15 & \multicolumn{1}{c|}{74060 ($\sim$ 20 hr) \footnote[4]}  & 9.753 & 7590x \footnotemark[2] \\ 
16 & \multicolumn{1}{c|}{1110900 ($\sim$ 308 hr) \footnotemark[4]} & 151.387 & 7340x \footnotemark[2] \\ 
17 & \multicolumn{1}{c|}{$\sim$ 4936 hr \footnotemark[4]} & 1867.08 & 9520x \footnotemark[2]\\ \hline
% 18 & \multicolumn{1}{c|}{5289.317} & 0.822 & 6427.022 \\ \hline
\end{tabular}
\caption{Execution times \& speedups achieved using CUDA for various values of $N$}
\label{tab:cuda}
\end{table}

\begin{figure}[!h]
    \centering
    \includegraphics[width=0.8\linewidth]{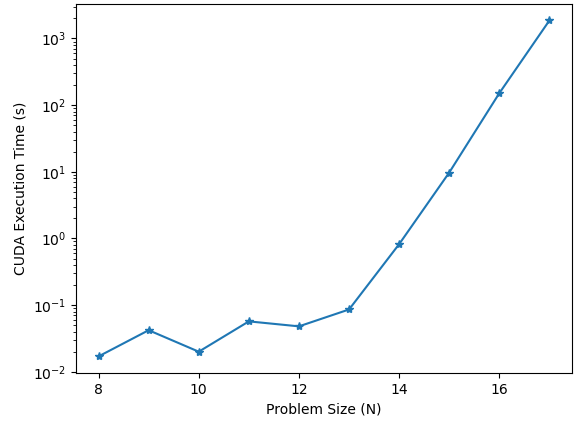}
    \caption{Plot of CUDA execution time w.r.t problem size ($N$)}
    \label{fig:cuda_time}
\end{figure}

\footnotetext[4]{Approximated using $O(N!)$ time trend of serial algorithm}
\footnotetext[2]{Estimated using approximated serial time \& actual CUDA time}

Figure \ref{fig:cuda_time} also shows the variation of CUDA execution time with increase in problem size $N$. We were restricted by the maximum capacity of C++ primitive datatypes (for storing the factorial values) to keep our analysis only upto $N=17$.

\balance

Some observations made from the above study include:

\begin{itemize}
    \item As $N$ increases, it is evident that the GPU provides immense benefits through parallelization, as we obtain speedups of great magnitudes.
    \item For large $N$ ($\geq 15$), it was infeasible to run the serial code due to the estimated total time based on the $O(N!)$ nature of the algorithm.
    \item For $N \geq 13$, even though the speedups are huge, the actual CUDA execution time increases exponentially because $N!$ becomes overwhelmingly large to be parallelized across $50 \times 1024$ threads.
    \item Following the trend, if $N = 18$, the serial execution could take around $83900$ hrs. Now, even if we extrapolate the speedups to $\sim 10000$x, the CUDA code will still require $8+$ hrs to execute, which is unacceptably large.
    \item \textbf{This analysis clearly illustrates the imperative need to choose an efficient algorithm to solve a problem instead of trying to parallelize an inefficient algorithm.}
\end{itemize}

\section{Conclusions \& Future Work}

In this project, we explore the Travelling Salesman Problem \& implement several approaches to parallelize the Brute Force TSP algorithm, including OpenMP, MPI \& CUDA programming. We perform detailed timing analysis for all the approaches \& also critically compare the performance of OpenMP \& MPI codes through their efficiency \& Karp-Flatt metric. We also develop and analyse a hybrid parallel algorithm leveraging both OpenMP \& MPI.

This project can be extended to parallelize \& compare other algorithms to solve the TSP such as Branch-\&-Bound, Genetic Algorithms, etc. This would provide a better idea about the relative importance of selecting a good algorithm \& parallelizing an algorithm.

\section*{References}

\scriptsize{
\begin{flushleft}
\begin{enumerate}
    \item \href{https://blog.routific.com/travelling-salesman-problem}{Understanding The Travelling Salesman Problem (TSP)}
    \item Burkhovetskiy, V. \& Steinberg, B.; \textit{Parallelizing an exact algorithm for the traveling salesman problem}; Procedia Computer Science, Volume 119, Pg 97-102 - 2017
    \item Izzatdin, Abdul \& Haron, Nazleeni \& Mehat, Mazlina \& Low, Tang \& Nabilah, Aisyah. (2008). \textit{Solving traveling salesman problem on high performance computing using message passing interface.}
    \item \href{http://www.mathcs.emory.edu/~cheung/Courses/355/Syllabus/94-CUDA/shared-vars.html}{Sharing global and local variables using Unified Memory in CUDA}
    \item \href{https://en.wikipedia.org/wiki/Karp\%E2\%80\%93Flatt\_metric}{Karp-Flatt Metric}
\end{enumerate}
\end{flushleft}
}

\end{document}